\documentclass[sigconf,nonacm]{acmart}

\AtBeginDocument{}

\usepackage{xspace}
\usepackage{colortbl}
\usepackage{amsmath,amsfonts}
\usepackage{algorithm}
\usepackage{algpseudocode}
\usepackage{graphicx}
\usepackage{textcomp}
\usepackage{xcolor}
\usepackage{tikz}
\usetikzlibrary{mindmap,shadows}
\usepackage[most]{tcolorbox}
\usepackage{graphbox}
\usepackage{pgfgantt}
\usepackage{mathtools}
\usepackage{fancybox}
\usepackage{makecell}
\usepackage{multirow}
\usepackage{booktabs}
\usepackage{hyperref}
\hypersetup{hidelinks}
\usepackage{cleveref}
\usepackage[normalem]{ulem} \usepackage{listings}
\usepackage{caption}
\usepackage{subcaption}
\usepackage{longtable}
\usepackage{pifont}\usepackage{enumitem}
\usepackage{subfloat}
\usepackage[latin1]{inputenx}
\usepackage[T1]{fontenc}
\usepackage{soul}
\usepackage{courier}
\usepackage{svg}

\newboolean{showcomments}
\setboolean{showcomments}{true}
\ifthenelse{\boolean{showcomments}}
{\newcommand{\nb}[2]{
    \fcolorbox{black}{yellow}{\bfseries\sffamily\scriptsize#1}
    {\sf\small$\blacktriangleright$\textit{#2}$\blacktriangleleft$}
    }
    
}
{\newcommand{\nb}[2]{}
    
}

\usepackage[framemethod=TikZ]{mdframed}

\newcommand\phase[1]{\tikz[baseline=(X.base)]\node [draw=black,fill=white,thick,rectangle,inner sep=2pt, rounded corners=2pt](X){\color{black}\textbf{#1}};}

\setcopyright{acmlicensed}
\copyrightyear{2026}
\acmYear{2026}
\acmDOI{XXXXXXX.XXXXXXX}
\acmConference[ICSE 2026]{48th International Conference on Software Engineering}{April 2026}{Rio de Janeiro, Brazil}
\acmISBN{978-1-4503-XXXX-X/2018/06}

\definecolor{light-gray}{rgb}{0.56, 0.74, 0.56}
\definecolor{light-blue}{rgb}{0.12, 0.53, 0.90}

\newcommand{\fitness}{\ensuremath{f}\xspace}
\newcommand{\asmpt}{\ensuremath{A}\xspace}
\newcommand{\budget}{\ensuremath{T}\xspace}
\newcommand{\requirement}{\ensuremath{\varphi}}

\newcommand{\system}{\ensuremath{S}}

\newcommand{\testsequence}{\textsc{TS}\xspace}
\newcommand{\failingtestsequence}{\textsc{TC}}
\newcommand{\candidatetestsequence}{\ensuremath{\testsequence_{c}}\xspace}

\newcommand{\testassessment}{\textsc{TA}\xspace}
\newcommand{\searchspace}{\textsc{SP}}
\newcommand{\fitnessfunction}{\textsc{FF}}
\newcommand{\fitnessvalue}{\textsc{f}}

\newcommand{\nff}{\textsc{NFF}\xspace}

\newcommand{\ATheNA}{\textsc{ATheNA}\xspace}
\newcommand{\HECATE}{\textsc{Hecate}\xspace}
\newcommand{\Athena}{\ATheNA}
\newcommand{\Hecate}{\HECATE}

\begin{document}

\title{Search-based Software Testing Driven by Domain Knowledge: Reflections and New Perspectives}

\author{Federico Formica}
\email{formicaf@mcmaster.ca}
\orcid{0000-0002-3033-7371}
\affiliation{\institution{McMaster University}
  \city{Hamilton}
  \state{ON}
  \country{Canada}
}

\author{Mark Lawford}
\email{lawford@mcmaster.ca}
\orcid{0000-0003-3161-2176}
\affiliation{\institution{McMaster University}
  \city{Hamilton}
  \state{ON}
  \country{Canada}
}

\author{Claudio Menghi}
\email{claudio.menghi@unibg.it}
\orcid{0000-0001-5303-8481}
\affiliation{\institution{University of Bergamo}
  \city{Bergamo}
  \state{BG}
  \country{Italy}
}
\affiliation{\institution{McMaster University}
  \city{Hamilton}
  \state{ON}
  \country{Canada}
}

\renewcommand{\shortauthors}{Formica et al.}

\begin{abstract}
    Search-based Software Testing (SBST) can automatically generate test cases to search for requirements violations. 
    Unlike manual test case development, it can generate a substantial number of test cases in a limited time.
    However, SBST does not possess the domain knowledge of engineers. 
    Several techniques have been proposed to integrate engineers' domain knowledge within existing SBST frameworks.
    This paper will reflect on recent experimental results by highlighting bold and unexpected results.
    It will help re-examine SBST techniques driven by domain knowledge from a new perspective, suggesting new directions for future research.
\end{abstract}

\begin{comment}
\begin{CCSXML}
<ccs2012>
   <concept>
       <concept_id>10011007</concept_id>
       <concept_desc>Software and its engineering</concept_desc>
       <concept_significance>500</concept_significance>
       </concept>
   <concept>
       <concept_id>10011007.10011074.10011099.10011102.10011103</concept_id>
       <concept_desc>Software and its engineering~Software testing and debugging</concept_desc>
       <concept_significance>500</concept_significance>
       </concept>
 </ccs2012>
\end{CCSXML}

\ccsdesc[500]{Software and its engineering}
\ccsdesc[500]{Software and its engineering~Software testing and debugging}
\end{comment}


\maketitle

\section{Introduction}
\label{sec:intro}

Cyber-Physical Systems (CPS) are composed of software and hardware components that communicate and interact.
CPS are used in many safety-critical industries, such as automotive \cite{matinnejad2015search}, aerospace \cite{Aristeo}, and biomedical \cite{majikes2013literature}. 
Within these domains, software failures can have catastrophic consequences.
For this reason, CPS models are extensively verified to avoid failures \cite{duan2018systematic}.

Testing is a technique used to identify software failures. 
CPS testing should not consider the software components in isolation, but also their interaction with their physical environment. 
This characteristic makes the testing activity more laborious, time-consuming, and expensive.
Indeed, to \emph{manually define} the test cases, engineers should consider both the behavior of the software and the reaction of its environment. 
The research and industrial communities are investing significant effort in improving testing strategies since they can prevent catastrophic events by identifying failures before the software is deployed on the CPS.
Therefore, improving SBST solutions is a relevant and widely recognized software engineering problem~\cite{papadakis2019mutation,5210118}.

Search-based Software Testing (SBST) is an iterative approach that has proven effective in dealing with CPS and identifying requirement violations.
SBST typically transforms the test case generation problem into an optimization problem. 
It tries to maximize or minimize a fitness function that guides the search toward the most critical test cases.
This approach offers one main advantage: The test cases are \emph{automatically generated} and are not manually developed by engineers.
Therefore, SBST can generate and assess a considerable number of test cases in a limited time.
However, SBST frameworks do not have the reasoning capabilities typical of engineers.

Engineers have \emph{extensive knowledge} of the operational context, requirements, and expected behavior of their CPS.
Classical SBST frameworks from the literature are not explicitly designed to effectively combine the domain knowledge of engineers to drive the search procedure~\cite{Taxonomy}. 
This capability can help identify software failures more effectively and efficiently.

For these reasons, recent SBST solutions (e.g.,~\cite{ATheNA,Hecate,RequirementsTables}) have incorporated engineers' domain knowledge within existing SBST frameworks.
They proposed frameworks that can combine human and machine capabilities to generate test cases more effectively.
Experimental results indicated that this synergic approach can outperform traditional SBST techniques. Additionally, it can find failure-revealing test cases that traditional SBST can not detect. 

This paper reflects on significant research contributions that propose \emph{SBST solutions driven by domain knowledge}.
It impacts the community by (a)~summarizing the contributions and their evaluation results, (b)~reflecting on the type of domain knowledge used to guide the search and evaluation practices, and (c)~proposing different types of domain knowledge artifacts that can be used to guide the search process. 

The use of \emph{domain knowledge} information for SBST may disrupt current SBST practices.
SBST solutions driven by domain knowledge are novel and innovative contributions. 
This paper provides bold and unexpected reflections that can help us reevaluate current research directions in a new light. For example, it provides additional arguments supporting the statement that ``\emph{generalizability is overrated}''~\cite{8048656}, and that the evaluation of SBST solutions should focus more on the quality of the findings instead of focusing on the quantity of study subjects considered in their evaluation.
Finally, this paper proposes new directions for future research by defining different types of domain knowledge artifacts that can be used to guide the search process.

We remark that domain knowledge have been considered within the testing community in several works (e.g.,~\cite{xue2024domain, Gambi_2022_Generating, Khatiri_2023_Simulation}). 
Unlike existing works, this paper advocates the use of domain knowledge as key ingredient to guide the SBST frameworks and provides reflections and additional types of domain knowledge that can be considered in future works.

This paper is organized as follows.
\Cref{sec:sbst} summarizes existing SBST solutions driven by domain knowledge.
\Cref{sec:results} summarizes evaluation results.
\Cref{sec:reflectionsandNew} re-examine SBST techniques from a new perspective, suggesting new directions
for future research. 
\Cref{sec:future} presents our future plans.
\Cref{sec:conclusion} concludes this work.

 \section{SBST Driven by Domain Knowledge}
\label{sec:sbst}
\Cref{fig:approach} provides a high-level overview of SBST driven by domain knowledge. 
Intuitively, the high-level idea is to enrich SBST with domain knowledge from engineers to perform a more effective and efficient search.
Domain knowledge can come from many sources and in different forms. In the following, we summarize some instantiations of SBST driven by domain knowledge.

\ATheNA~\cite{ATheNA,ATheNA_Tool}  embeds domain knowledge within SBST fitness functions.
\Cref{fig:athena} provides an overview of \ATheNA.
\ATheNA takes as input a system (\system) and some assumptions on its inputs ($A$).
It returns a failure-revealing test case ($TC$) or an indication (NFF --- no failure found) that no failure was found within the time budget.
\ATheNA combines (step \phase{2}) fitness functions automatically generated and manually-defined by engineers to generate new test cases ($TC$).
Automatically generated fitness functions ($f_a$ --- step \phase{4}) typically come from the system requirements ($\requirement$).
They are effective in many practical cases.
However, engineers possess additional knowledge about conditions that may lead to system failures. 
For example, if a requirement specifies that the speed of a vehicle should not exceed $120$mph, increasing the throttle may raise the likelihood of violating this requirement.
\ATheNA enables engineers to embed this domain knowledge within the search process. 
Specifically, this information (step \phase{5}) can be embedded into a manual fitness function ($f_m$), which is combined with the automatic fitness (\fitness --- step \phase{3}).
This framework enables the SBST to use information about the system requirements and domain knowledge to drive the search (step \phase{1}) for failure-revealing test cases.

\ATheNA is implemented within a tool~\cite{ATheNA_Tool} that supports an instance of the general approach targeting Simulink models.
\ATheNA is available on GitHub~\cite{ATheNAGitHub} and on Matlab File Exchange (Matlab marketplace)~\cite{ATheNAMatlab}.

\tikzstyle{output} = [coordinate]
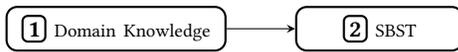
\begin{figure}[t]
    \centering
    \begin{tikzpicture}[auto,
 block/.style ={rectangle, draw=black, thick, fill=white!20, text width=5em,align=center, rounded corners},
 block1/.style ={rectangle, draw=blue, thick, fill=blue!20, text width=5em,align=center, rounded corners, minimum height=2em},
 line/.style ={draw, thick, -latex',shorten >=2pt},
 cloud/.style ={draw=red, thick, ellipse,fill=red!20,
 minimum height=1em}]

\node [block,node distance=2.2cm,text width=2.7cm] (FFG) {\phase{1} \footnotesize Domain Knowledge};
\node [block,right of=FFG,node distance=3.5cm,text width=2.0cm] (SE) {\phase{2} \footnotesize SBST};
\node [output, right of=SE,node distance=2.5cm] (OUT) {};

\draw[-stealth] (FFG.east) -- (SE.west)    node[pos=0.5, above]{};
\end{tikzpicture}     \caption{SBST Driven by Domain Knowledge.}
    \label{fig:approach}
    \Description[Flowchart for Domain Knowledge input into SBST.]{
    The figure contains a flowchart with two blocks. The first block contains the Domain Knowledge, which is an input to the second block, which contains the Search-based Software Testing algorithm.
    }
\end{figure}

\tikzstyle{output} = [coordinate]
\begin{figure}[t]
    \centering
    \begin{tikzpicture}[auto,
 block/.style ={rectangle, draw=black, thick, fill=white!20, text width=5em,align=center, rounded corners},
 block1/.style ={rectangle, draw=blue, thick, fill=blue!20, text width=5em,align=center, rounded corners, minimum height=2em},
 line/.style ={draw, thick, -latex',shorten >=2pt},
 cloud/.style ={draw=red, thick, ellipse,fill=red!20,
 minimum height=1em}]

\draw (0,0) node[block] (Input) {\phase{1} \footnotesize Search};
\node [block, right of=Input, node distance=4cm, minimum height=3.5cm, minimum width=4.5cm] (OverallFitness) {};
\node [block, above of=OverallFitness, node distance=1.3cm, line width=0mm, draw=white!80, text width=4.1cm] (OverallFitnessLabel) {\phantom{xxxxxxx} \phase{2} \footnotesize Fitness Assessment};
\node [output, above of=OverallFitness, node distance=0.5cm] (ftn) {};
\node[block, below of=ftn, node distance=1.3cm] (Requirement){\phase{3} \footnotesize \Athena\\ Fitness
};
\node [block, right of=ftn, node distance=1.1cm] (ManualFitness) {\phase{5} \footnotesize Manual\\  Fitness};
\node [block, left of=ftn, node distance=1.1cm] (AutomaticFitness) {\phase{4} \footnotesize Automatic\\ Fitness};
\node [output, right of=OverallFitness, node distance=4cm] (uoutput) {};
\node [output, above of=Input, node distance=3mm] (Asmpt1) {};
\node [output, left of=Asmpt1, node distance=4mm] (Asmpt2) {};
\node [output, above of=Asmpt2, node distance=6mm] (Constraints) {};
\node [output, right of=Asmpt1, node distance=4mm] (InputModel2) {};
\node [output, above of=InputModel2, node distance=6mm] (InputModel) {};
\node [output, below of=Requirement, node distance=0.7cm] (v1a) {};
\node [output, left of=v1a, node distance=3cm] (v1) {};
\node [output, left of=v1, node distance=1cm] (v2) {};
\node [output, below of=ManualFitness, node distance=1.2cm] (mftwo) {};
\node [output, below of=AutomaticFitness, node distance=1.2cm] (mf) {};
\node [output, above of=AutomaticFitness, node distance=1cm] (ff) {};
\node [output, left of=OverallFitness, node distance=2.25cm] (budget1) {};
\node [output, above of=budget1, node distance=1cm] (budget2) {};
\node [output, left of=budget2, node distance=0.7cm] (budget3) {};

\draw[-stealth] (budget3) -- (budget2)
    node[midway,above]{$\budget$};
\draw[-stealth] (Constraints.south) -- (Asmpt2.north)
    node[midway,left]{$\asmpt$};
\draw[-stealth] (InputModel.south) -- (InputModel2.north)
    node[midway,right]{$\system$};
\draw[-] (v1) -- (v2)
    node[midway,below]{$\fitness$};
\draw[-stealth] (v2) -- (Input.south);
\draw[-stealth] (Input) -- (OverallFitness.west)
    node[midway,above]{$TC$};
\draw[-stealth] (OverallFitness.east) -- (uoutput)
    node[midway,above]{$TC$/NFF};
\draw[-] (AutomaticFitness.south) -- (mf)
    node[midway,left]{$\fitness_a$};
\draw[-stealth] (mf) -- (Requirement)
    node[midway,left]{};
\draw[-] (ManualFitness.south) -- (mftwo)
    node[midway,right]{$\fitness_m$};
\draw[-stealth] (mftwo) -- (Requirement)
    node[midway,left]{};
\draw[-stealth] (ff) -- (AutomaticFitness)
    node[midway,left]{$\requirement$};
\draw[-] (Requirement) -- (v1a)
    node[midway,left]{};
\draw[-] (v1a) -- (v1)
    node[midway,left]{};

\end{tikzpicture}
    \caption{Integrating Domain Knowledge within SBST by exploiting Fitness Functions.}
    \label{fig:athena}
    \Description[Flowchart for the \Athena framework.]{
    The figure contains a flowchart with five blocks divided into two parts.
    In the part on the right, the Fitness Assessment block is refined into three separate blocks: \Athena fitness, Automatic fitness, and Manual fitness.
    On the left, there is a single block representing the search algorithm.
    The Fitness Assessment block and the Search block are connected in a feedback loop.
    }
\end{figure}
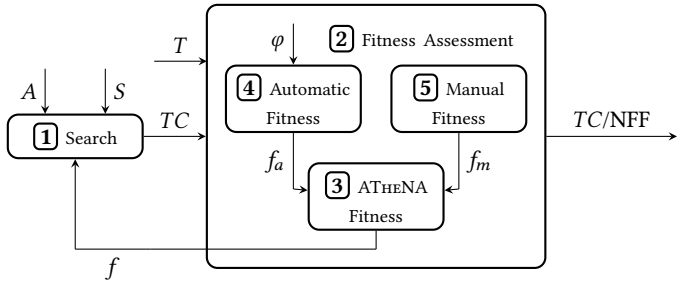

\HECATE~\cite{Hecate} uses the domain knowledge that engineers provided in previously defined test cases to generate failure-revealing test cases.
\Cref{fig:hecate} provides an overview of \Hecate. 
Specifically, \HECATE receives as input a test case composed of a test sequence (TS) and a test assessment (TA) block.
The test sequence block specifies the inputs that are considered by the test case. 
The test assessment block specifies the oracle that determines if the test is passed or not.  
\HECATE returns a failure-revealing test case ($TC$) or an indication (NFF --- no failure found) that no failure was found within the time budget.
\HECATE is composed by two phases:  \emph{driver} and \emph{search}.
The \emph{driver} phase translates the 
test case into artifacts that drive the simulation-based search.
First, the test sequence previously defined by the engineers is extended to encode the search space (step \phase{1}).
Specifically, \HECATE enables engineers to inject search parameters within the test sequence block, leading to a parameterized test sequence (\searchspace). 
This procedure enables engineers to reuse domain knowledge from the test sequence blocks and use this information within the search procedure, thanks to the search parameters. 
\HECATE translates the Test Assessment block into a fitness function (\fitnessfunction) that guides the search (Step \phase{2}). 
The \emph{search} phase implements the iterative testing procedure of the simulation-based software testing framework in two steps: It iteratively generates (step~\phase{3}) new candidate Test Sequences (\candidatetestsequence) using the search space definition (\searchspace),  executes the system (\system) for the test inputs specified by the candidate Test Sequence, and checks (step~\phase{4}) using the fitness function (\fitnessfunction) whether the Test Assessment is satisfied or violated. \HECATE~\cite{RequirementsTables} was recently extended and can now also support the domain knowledge engineers provide within Tabular Requirements to guide the search process, and specifically Simulink Requirements Tables~\cite{menghi2024completeness}. 
Unlike existing approaches, this solution is the first to use requirements expressed in tabular form to generate fitness functions that guide the generation of test cases.

\HECATE was implemented within a tool that targets Simulink models. 
The implementation is available on GitHub \cite{HecateGitHub} and the Matlab File Exchange \cite{HecateMatlab}.

In summary, SBST techniques try to incorporate domain knowledge within the search process.
For example, in this section we presented \ATheNA and \HECATE, two search-based software testing tools that incorporated domain knowledge within the fitness function and the search domain.
The next section summarizes their evaluation and results.

\tikzstyle{output} = [coordinate]
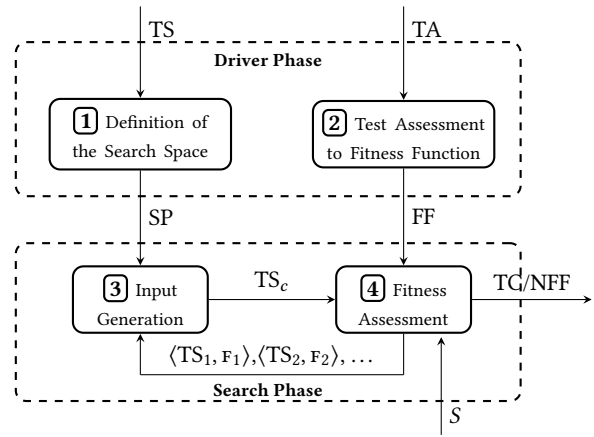
\begin{figure}[t]
    \centering
    \begin{tikzpicture}[auto,
 block/.style ={rectangle, draw=black, thick, fill=white!20, text width=5em,align=center, rounded corners},
 block1/.style ={rectangle, draw=blue, thick, fill=blue!20, text width=5em,align=center, rounded corners, minimum height=2em},
 line/.style ={draw, thick, -latex',shorten >=2pt},
 cloud/.style ={draw=red, thick, ellipse,fill=red!20,
 minimum height=1em}]

\draw (0,0) node[block] (Input) {\phase{3} \footnotesize Input \\ Generation};
\node [output, right of=Input,node distance=1.7cm] (InputMiddle) {};
\node [block, below of=InputMiddle,node distance=0.3cm,text width=6.5cm,minimum height=2.1cm,dashed,draw opacity=1] (Search) {};
\draw (0,0) node[block] (InputRef) {\phase{3} \footnotesize Input \\ Generation};
\node [block, above of=InputMiddle,node distance=2.4cm,text width=6.5cm,minimum height=2.05cm,dashed] (Driver) {};
\node[block, right of=Input,node distance=3.5cm] (Fitness){\phase{4} \footnotesize Fitness\\ Assessment};
\node [output, right of=Fitness,node distance=2.5cm] (uoutput) {};
\node [block, above of=Fitness,node distance=2.2cm,text width=2.2cm] (TestAssessment) {\phase{2} \footnotesize Test Assessment to Fitness Function};
\node [block, above of=Input,node distance=2.2cm,text width=2.2cm] (TestToInput) {\phase{1} \footnotesize Definition of the Search Space};
\node [output, above of=TestToInput,node distance=1.7cm] (testsequence) {};
\node [output, above of=TestAssessment,node distance=1.7cm] (testassessment) {};
\node [output, below of=Fitness,node distance=1cm] (loopa) {};
\node [output, below of=Input,node distance=1cm] (loopb) {};
\draw (0,0) node [above of=Driver, node distance=0.8cm] (DriverText) {\footnotesize \textbf{Driver Phase}};
\draw (0,0) node [below of=Search, node distance=0.9cm] (SearchText) {\footnotesize \textbf{Search Phase}};
\node [output, right of=Fitness,node distance=0.5cm] (modela) {};
\node [output, below of=modela,node distance=1.8cm] (model) {};
\node [output, below of=modela,node distance=0.5cm] (modelb) {};

\draw[-stealth] (model.east) -- (modelb.south) 
    node[pos=0.2,
right]{\system};
\draw[-stealth] (Input.east) -- (Fitness.west)
    node[midway,above]{\candidatetestsequence};
\draw[-stealth] (testsequence.south) -- (TestToInput.north)
    node[near start,right]{\testsequence};
\draw[-stealth] (TestToInput.south) -- (Input.north)
    node[midway,right]{\searchspace};
\draw[-stealth] (testassessment.south) -- (TestAssessment.north)
    node[near start,right]{\testassessment};
\draw[-stealth] (TestAssessment.south) -- (Fitness.north)
    node[midway,right]{\fitnessfunction};    
\draw[-] (Fitness.south) -- (loopa.north)
    node[midway,right]{};    
\draw[-] (loopb.south) -- (loopa.north)
    node[midway,above]{$\langle \testsequence_1, \fitnessvalue_1 \rangle$,$\langle \testsequence_2, \fitnessvalue_2 \rangle$, \ldots};
\draw[-stealth] (loopb.north) -- (Input.south)
    node[midway,right]{};    
\draw[-stealth] (Fitness.east) -- (uoutput.west)
    node[midway,above]{\failingtestsequence/\nff};

\end{tikzpicture}
    \caption{Integrating Domain Knowledge within SBST by exploiting existing test cases.}
    \label{fig:hecate}
    \Description[Flowchart for the \Hecate framework.]{
    The figure contains a flowchart with four blocks divided into two parts.
    In the group above, there are two blocks which form the Driver Phase: the Search Space Definition and the Test Assessment to Fitness Function generation.
    These two blocks are executed in parallel, and the output is fed to the blocks below.
    In the group below, there are two blocks forming the Search Phase: the Input Generation and the Fitness Assessment.
    These two blocks are connected in a loop and take as input the output of the Driver Phase.
    }
\end{figure}

\section{Evaluation and Results}
\label{sec:results}
We summarize the evaluation and results obtained for each SBST framework driven by domain knowledge described in \Cref{sec:sbst}.

\ATheNA was evaluated by considering seven benchmarks  from the ARCH competition~\cite{DBLP:conf/arch/ErnstABCDFFG0KM21} --- an international competition
among testing tools for continuous and hybrid systems~\cite{ARCHWEBSITE}.
The results confirm that (a)~it is practical for engineers to write effective fitness functions encoding their domain knowledge, (b)~the balance between fitness information coming from the requirement and the domain knowledge influence the effectiveness of the SBST procedure, and (c)~using the domain knowledge improves the effectiveness of the SBST procedure in most of the cases and does not introduce significant performance overheads. 
Additionally, \ATheNA was applied on two representative case studies: One from the automotive domain and one from the medical domain.  
\ATheNA successfully returned a failure-revealing test case for these case studies.
\ATheNA also participated in three editions of the ARCH competition~\cite{Arch2023, Arch2024,Arch2025}.
In all years, \ATheNA produced remarkable results, confirming its effectiveness.

\HECATE~\cite{Hecate}
was evaluated by considering 18 Simulink models, including models from the ARCH competition, from the web (\cite{simulinkpacemaker,TestSequence, benchmarkHPS,benchmarkST,benchmarkFS,benchmarkTL}), and from an industrial benchmark \cite{mavridou2020ten}.
The benchmark includes models developed by Toyota~\cite{jin2014powertrain} and Lockheed-Martin~\cite{mavridou2020ten,benchmarkLM}.
The results show that for the benchmark models, \HECATE is effective and efficient in finding failure-revealing test cases.
To evaluate the applicability of \Hecate, the authors selected a large and representative automotive case study developed by MathWorks in the context of the EcoCAR Mobility Challenge~\cite{EcoCAR}, a competition they jointly organize with the U.S.\ Department of Energy and General Motors.
They checked whether \Hecate could generate a failure-revealing test case.
\Hecate successfully returned a failure-revealing test case for the case study. 
\HECATE was also used to support the development of a Cruise Control (CC) model for an Industrial Simulator \cite{Hecate_CruiseControl}.
The tool supported the design of seven model versions, including 21 minor versions.
\HECATE effectively found faults and helped engineers improve their models. 
The version of \HECATE that supports Requirements Tables also showed remarkable effectiveness and efficiency, returning a failure-revealing test case for 85\% of the considered benchmarks.
Remarkably, it identified a failure-revealing test case for three versions of the CC model, which the previous version could not find.

 \section{Reflections and New Research Directions}
\label{sec:reflectionsandNew} 
We reflect on the existing results for SBST Driven by Domain Knowledge (\Cref{sec:reflections}), 
and present new research directions(\Cref{sec:directions}).

\subsection{Reflections}
\label{sec:reflections}
We reflect on the type of domain knowledge considered by the existing SBST driven by domain knowledge from \Cref{sec:sbst} and reflect on their evaluation.

\emph{Type of Domain Knowledge}. \ATheNA and \HECATE consider domain knowledge embedded within the fitness function and the test sequence.
They provided a paradigm shift for SBST. 
\ATheNA enables engineers to explicitly use their domain knowledge to help SBST define \emph{how} to search for failure-revealing test cases.
\HECATE enables engineers to use the domain knowledge encoded in existing test cases to help SBST define \emph{where} to search for failure-revealing test cases.
It also supports the use of information coming from Test Assessment and Requirements Tables to help SBST define \emph{why} to search for failure-revealing test cases.
However, SBST driven by domain knowledge (\Cref{fig:approach}) can be extended to use information from many different sources, e.g., information related to regression testing. 
It can also be extended to support the generation of different types of test cases (\emph{what}), that are needed by different users using the SBST solution~(\emph{who}).

\emph{Evaluation}.
SBST driven by domain knowledge is implicitly context-driven. 
While \ATheNA and \HECATE are general solutions that can be applied to solve different problems related to many systems, domain knowledge typically refers to a precise context. 
For example, the domain knowledge considered for supporting the development of the Cruise Control (CC) model for the VI-CarRealTime industrial simulator~\cite{Hecate_CruiseControl} comes from test cases developed for this system, which represent an external threat to validity for the results of this work.
The importance of the context in software engineering solutions has also been discussed by Briand et al.~\cite{8048656}, who advocated the need for context-driven research.
The authors argue that the background of engineers and domain-related elements are some of the factors that influence the applicability and scalability of research.
To mitigate this problem, \ATheNA and \HECATE considered a large set of benchmarks from different domains:
automotive,     aerospace,      medical,        energy,         home appliances, and defense.    They also considered models from 
General Motors,     Toyota,             Lockheed Martin,    MathWorks,          and VI-Grade.       For example, for one of the benchmarks from the automotive domain, \HECATE returned a significantly lower number of failure-revealing test cases compared to the other benchmarks.
However, the performance reduction was not caused by the \HECATE framework, but by the domain knowledge and the system considered in this experiment.
Considering these arguments, we support the claim from Briand et al~\cite{8048656} that ``\emph{generalizability is overrated}'' in the context of the assessment of SBST solutions driven by domain knowledge.

We also remark that considering such a variety of case studies is complex and time-consuming.
While many industries are sharing their code, software companies typically do not share their Simulink models~\cite{boll2024replicability,boll2020replicability,Boll_2021_Characteristics}.
Therefore, assessing SBST solutions driven by domain knowledge on models (e.g., Simulink models) introduces additional challenges due to the limited number of publicly available benchmarks.
For example, the development and testing of the CC model required a considerable time (eight months~\cite{Hecate_CruiseControl}).
A recent study from the e-Bike domain mentioned that the development of testing activities required 100 days (approximately three months)~\cite{marzella2025test}.
This time does not include the time necessary to study and understand the application domain, which brings it to more than one year.
Indeed, to understand the domain, results, and assess the SBST framework, there is a need to significantly interact with domain experts who have the know-how necessary to provide valuable and significant interpretation to the research results.
Therefore, evaluating SBST solutions driven by domain knowledge within the context of Simulink models may require years to consider multiple study subjects.
Considering this argument, we believe that the academic community should focus more on the quality of the evaluation and its findings rather than on the number of study subjects.
Specifically, qualitative results (on single or a limited number of benchmarks) rigorously discussed and analyzed with domain experts are more valuable than numerical results that have limitations due to their non-generalizability and domain-specificity.

\subsection{New Research Directions}
\label{sec:directions}
\Cref{fig:taxonomy} provides a (non-exhaustive) set of domain knowledge information that can guide SBST solutions. 
It summarizes sources of domain knowledge that have already been considered (blue nodes), and new sources of domain knowledge (green nodes).

\noindent $\bullet$ \emph{Requirements} - Domain knowledge coming from requirements has been considered to guide the SBST solution (e.g.,~\cite{Aristeo, matinnejad2015search}). 

\noindent $\bullet$ \emph{Search Algorithm} - Domain knowledge coming from experts can guide the choice of the search algorithm. 
It can be used to select between random and guided search, to define the mutation operators for the genetic algorithm, and configure the algorithms to favor exploration vs exploitation. 
For example, recent studies~\cite{RequirementsTables,marzella2025test} show complementary results for uniform random and simulated annealing:  
Unlike random solutions, non-random solutions (e.g., SA) are more effective when failures are difficult to find since the use of a fitness metric to guide the search process is beneficial.

\noindent $\bullet$ \emph{Fitness Function} - Domain knowledge has been encoded within the fitness function to help prioritize the generation of test cases with specific characteristics (e.g.,~\cite{ATheNA,ATheNA_Tool}).

\noindent $\bullet$ \emph{Manually Defined Test Cases} - Domain knowledge coming from previously defined test cases have been considered to guide the SBST solution (e.g.,~\cite{Hecate}).

\noindent $\bullet$ \emph{System Documentation} - Information coming from system documentation can be considered for generating new test cases.
System documentation can include additional information (which is not necessarily part of the requirements). 
For example, knowing that a system has been implemented by considering specific design patterns may help generate test cases tailored to these patterns and trigger recurrent critical situations. 

\noindent $\bullet$ \emph{Bug Reports} - Domain knowledge coming from bug reports can help SBST solutions prioritize the generation of test cases that are more likely to trigger these bugs.
For example, when designing vehicle controllers, bugs detected in similar controllers may be used to generate test cases that are more likely to trigger them.

\noindent $\bullet$ \emph{Outcome of the Test Cases} - Domain knowledge coming from the execution of previous test cases can help SBST solutions generate more critical test cases.
For example, a test case that has failed frequently is more critical than one that always passes. 
This information can be used to guide the SBST to generate test cases that are more similar to those that failed more frequently. 

 \noindent $\bullet$ \emph{Development Information} - Domain knowledge coming from the development can help SBST generate more critical test cases. 
 For example, if a portion of a model (or source code) is changed more frequently, it is more likely to contain bugs and problems. 
 This information can be used to guide the SBST to generate test cases that focus on these parts of the model (or source code). 

\begin{figure}[t]
    \centering

\begin{tikzpicture}[every annotation/.style = {draw, fill = black, font = \scriptsize}]
    \path[mindmap, concept color=light-gray, text=black,
        every node/.style = {concept, circular drop shadow, font=\scriptsize},
        root/.style = {concept color=light-gray, font=\scriptsize, text width=6em, minimum size=6em},
        level 1 concept/.append style = {font=\scriptsize, sibling angle=45, text width=6em, level distance=9em, inner sep=0pt, minimum size=6em},
        level 2 concept/.append style = {font=\scriptsize, level distance=7em, sibling angle=37, minimum size=3.2em},
    ]
    node[root] {Domain\\ Knowledge} 
        [clockwise from=90]
    child[concept color=light-blue] {
        node[concept] {Requirements}
    }
    child[concept color=light-blue] {
        node {Search Algorithm}
    }
    child[concept color=light-blue] {
        node {Fitness Function}
    }
    child[concept color=light-blue] {
        node[concept] {Manual Test Cases}
    }
    child[concept color=light-gray]{
        node[concept] {System\\ Documentation}
    }
    child[concept color=light-gray] {
        node[concept] (TeXdoc) {Bug Reports}
    }
    child[concept color=light-gray] {
        node[concept] (TeXdoc) {Outcome of the Test Cases}
    }
    child[concept color=light-gray] {
        node[concept] (Blogs) {Development Information}
    };
\end{tikzpicture}
    \caption{Type of domain knowledge.}
    \label{fig:taxonomy}
    \Description[Conceptual map for the types of domain knowledge.]{
    Conceptual map structured like a wheel.
    In the middle, there is a bubble with the words ``Domain Knowledge''.
    Surrounding it at 45-degree angles, there are the following eight concepts in clockwise order: Purpose, Usage, Subject, Design, Coverage, Documentation, Execution, and Test Cases.
    }
\end{figure}
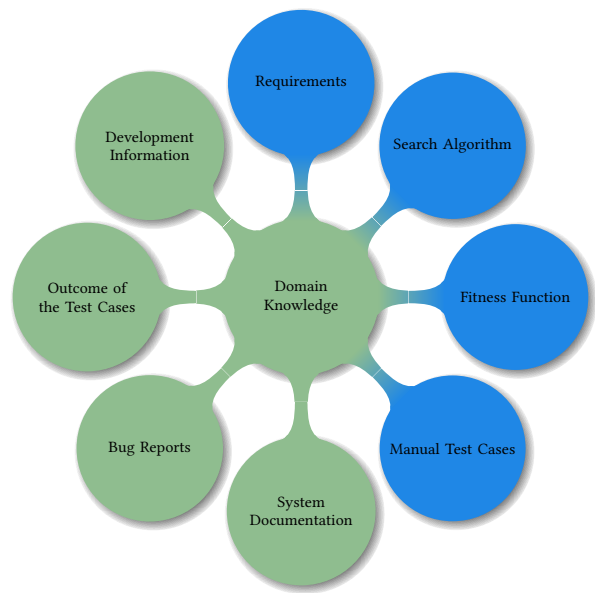 \section{Future Plans}
\label{sec:future}

\begin{figure}[t]
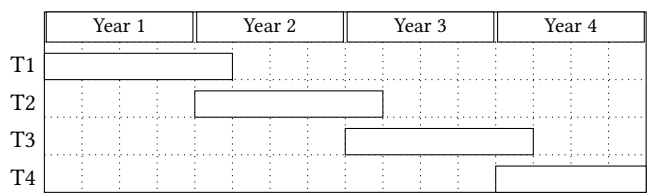

    \begin{center}
    
    \begin{ganttchart}[y unit title=0.4cm,
    y unit chart=0.5cm,
    vgrid,hgrid, 
    title label anchor/.style={below=-1.6ex},
    title left shift=.05,
    title right shift=-.05,
    title height=1,
    progress label text={},
    bar height=0.7,
    group right shift=0,
    group top shift=.6,
    group height=.3]{1}{16}
\gantttitle{Year 1}{4} 
    \gantttitle{Year 2}{4} 
    \gantttitle{Year 3}{4} 
    \gantttitle{Year 4}{4}  \\
\ganttbar{T1}{1}{5} \\
    \ganttbar{T2}{5}{9} \\
    \ganttbar{T3}{9}{13} \\
    \ganttbar{T4}{13}{16}

\end{ganttchart}
    \end{center}
    \caption{Gantt Chart}
    \label{fig:Gant}
\end{figure}

Incorporating domain knowledge within SBST solutions is a community process where the community identifies sources of domain knowledge and embedded them within the search process. 
This paper advocates this need, summarizes the results obtained for by considering four types of domain knowledge information and proposes and identified four additional sources of domain knowledge that can be incorporated within the search procedure. 

The Gantt Chart from \Cref{fig:Gant} summarizes the future plans for incorporating the type of domain knowledge from \Cref{sec:directions} within SBST solutions. 
Tasks T1, T2, T3, and T4 refer to \emph{System Documentation}, \emph{Bug Reports}, \emph{Outcome of the Test Cases}, and \emph{Development Information}. 
We allocate one year for incorporating each type of domain knowedge within existing SBST solutions. \section{Conclusion}
\label{sec:conclusion}

This paper presented original reflections and directions for future research on SBST driven by domain knowledge.
To ensure the soundness of our reflections and new research directions, we referred to technical and theoretical contributions reported in previous publications (e.g.,~\cite{ATheNA,ATheNA_Tool,Hecate}).

\bibliographystyle{ACM-Reference-Format}

\begin{thebibliography}{40}



\ifx \showCODEN    \undefined \def \showCODEN     #1{\unskip}     \fi
\ifx \showISBNx    \undefined \def \showISBNx     #1{\unskip}     \fi
\ifx \showISBNxiii \undefined \def \showISBNxiii  #1{\unskip}     \fi
\ifx \showISSN     \undefined \def \showISSN      #1{\unskip}     \fi
\ifx \showLCCN     \undefined \def \showLCCN      #1{\unskip}     \fi
\ifx \shownote     \undefined \def \shownote      #1{#1}          \fi
\ifx \showarticletitle \undefined \def \showarticletitle #1{#1}   \fi
\ifx \showURL      \undefined \def \showURL       {\relax}        \fi
\providecommand\bibfield[2]{#2}
\providecommand\bibinfo[2]{#2}
\providecommand\natexlab[1]{#1}
\providecommand\showeprint[2][]{arXiv:#2}

\bibitem[Ali et~al\mbox{.}(2010)]{5210118}
\bibfield{author}{\bibinfo{person}{Shaukat Ali}, \bibinfo{person}{Lionel~C.
  Briand}, \bibinfo{person}{Hadi Hemmati}, {and}
  \bibinfo{person}{Rajwinder~Kaur Panesar-Walawege}.}
  \bibinfo{year}{2010}\natexlab{}.
\newblock \showarticletitle{A Systematic Review of the Application and
  Empirical Investigation of Search-Based Test Case Generation}.
\newblock \bibinfo{journal}{\emph{IEEE Transactions on Software Engineering}}
  \bibinfo{volume}{36}, \bibinfo{number}{6} (\bibinfo{year}{2010}),
  \bibinfo{pages}{742--762}.
\newblock
\href{https://doi.org/10.1109/TSE.2009.52}{doi:\nolinkurl{10.1109/TSE.2009.52}}


\bibitem[ARCH(line)]{ARCHWEBSITE}
ARCH organizational committee \bibinfo{year}{2021 [Online]}\natexlab{}.
\newblock \bibinfo{booktitle}{\emph{{International Competition on Verifying
  Continuous and Hybrid Systems}}}.
\newblock ARCH organizational committee.
\newblock
\urldef\tempurl \url{https://cps-vo.org/group/ARCH/FriendlyCompetition}
\showURL{Retrieved April 2022 from \tempurl}


\bibitem[AVTC(2022)]{EcoCAR}
\bibfield{author}{\bibinfo{person}{AVTC}.} \bibinfo{year}{2022}\natexlab{}.
\newblock \bibinfo{title}{{The EcoCAR Mobility Challenge}}.
\newblock
  \bibinfo{howpublished}{\url{https://avtcseries.org/about-avtc/past-competitions/ecocar-mobility-challenge/}}.
\newblock


\bibitem[Ayesh et~al\mbox{.}(2022)]{simulinkpacemaker}
\bibfield{author}{\bibinfo{person}{Mostafa Ayesh}, \bibinfo{person}{Namya
  Mehan}, \bibinfo{person}{Ethan Dhanraj}, \bibinfo{person}{Abdul El-Rahwan},
  \bibinfo{person}{Simon~Emil Opalka}, \bibinfo{person}{Tony Fan},
  \bibinfo{person}{Akil Hamilton}, \bibinfo{person}{Akshay~Mathews Jacob},
  \bibinfo{person}{Rahul~Anthony Sundarrajan}, \bibinfo{person}{Bryan Widjaja},
  {and} \bibinfo{person}{Claudio Menghi}.} \bibinfo{year}{2022}\natexlab{}.
\newblock \showarticletitle{Two Simulink Models with Requirements for a Simple
  Controller of a Pacemaker Device}. In \bibinfo{booktitle}{\emph{International
  Workshop on Applied Verification of Continuous and Hybrid Systems (ARCH22)}}
  \emph{(\bibinfo{series}{EPiC Series in Computing},
  Vol.~\bibinfo{volume}{90})}. \bibinfo{publisher}{EasyChair},
  \bibinfo{pages}{18--25}.
\newblock


\bibitem[Boll et~al\mbox{.}(2021)]{Boll_2021_Characteristics}
\bibfield{author}{\bibinfo{person}{Alexander Boll}, \bibinfo{person}{Florian
  Brokhausen}, \bibinfo{person}{Tiago Amorim}, \bibinfo{person}{Timo Kehrer},
  {and} \bibinfo{person}{Andreas Vogelsang}.} \bibinfo{year}{2021}\natexlab{}.
\newblock \showarticletitle{Characteristics, potentials, and limitations of
  open-source Simulink projects for empirical research}.
\newblock \bibinfo{journal}{\emph{Software and Systems Modeling}}
  \bibinfo{volume}{20}, \bibinfo{number}{6} (\bibinfo{year}{2021}),
  \bibinfo{pages}{2111--2130}.
\newblock
\href{https://doi.org/10.1007/s10270-021-00883-0}{doi:\nolinkurl{10.1007/s10270-021-00883-0}}


\bibitem[Boll and Kehrer(2020)]{boll2020replicability}
\bibfield{author}{\bibinfo{person}{Alexander Boll} {and} \bibinfo{person}{Timo
  Kehrer}.} \bibinfo{year}{2020}\natexlab{}.
\newblock \showarticletitle{On the Replicability of Experimental Tool
  Evaluations in Model-Based Development: Lessons Learnt from a Systematic
  Literature Review Focusing on MATLAB/Simulink}. In
  \bibinfo{booktitle}{\emph{International Conference on Systems Modelling and
  Management}}. Springer, \bibinfo{pages}{111--130}.
\newblock


\bibitem[Boll et~al\mbox{.}(2024)]{boll2024replicability}
\bibfield{author}{\bibinfo{person}{Alexander Boll}, \bibinfo{person}{Nicole
  Vieregg}, {and} \bibinfo{person}{Timo Kehrer}.}
  \bibinfo{year}{2024}\natexlab{}.
\newblock \showarticletitle{Replicability of experimental tool evaluations in
  model-based software and systems engineering with MATLAB/Simulink}.
\newblock \bibinfo{journal}{\emph{Innovations in Systems and Software
  Engineering}} \bibinfo{volume}{20}, \bibinfo{number}{3}
  (\bibinfo{year}{2024}).
\newblock


\bibitem[Briand et~al\mbox{.}(2017)]{8048656}
\bibfield{author}{\bibinfo{person}{Lionel Briand}, \bibinfo{person}{Domenico
  Bianculli}, \bibinfo{person}{Shiva Nejati}, \bibinfo{person}{Fabrizio
  Pastore}, {and} \bibinfo{person}{Mehrdad Sabetzadeh}.}
  \bibinfo{year}{2017}\natexlab{}.
\newblock \showarticletitle{The Case for Context-Driven Software Engineering
  Research: Generalizability Is Overrated}.
\newblock \bibinfo{journal}{\emph{IEEE Software}} \bibinfo{volume}{34},
  \bibinfo{number}{5} (\bibinfo{year}{2017}), \bibinfo{pages}{72--75}.
\newblock
\href{https://doi.org/10.1109/MS.2017.3571562}{doi:\nolinkurl{10.1109/MS.2017.3571562}}


\bibitem[Duan et~al\mbox{.}(2018)]{duan2018systematic}
\bibfield{author}{\bibinfo{person}{Pengfei Duan}, \bibinfo{person}{Ying Zhou},
  \bibinfo{person}{Xufang Gong}, {and} \bibinfo{person}{Bixin Li}.}
  \bibinfo{year}{2018}\natexlab{}.
\newblock \showarticletitle{A systematic mapping study on the verification of
  cyber-physical systems}.
\newblock \bibinfo{journal}{\emph{IEEE Access}}  \bibinfo{volume}{6}
  (\bibinfo{year}{2018}), \bibinfo{pages}{59043--59064}.
\newblock
Issue 2018.
\href{https://doi.org/10.1109/ACCESS.2018.2872015}{doi:\nolinkurl{10.1109/ACCESS.2018.2872015}}


\bibitem[Elliott et~al\mbox{.}(2015)]{benchmarkLM}
\bibfield{author}{\bibinfo{person}{Christopher Elliott},
  \bibinfo{person}{Gregory Tallant}, {and} \bibinfo{person}{Peter Stanfill}.}
  \bibinfo{year}{2015}\natexlab{}.
\newblock \showarticletitle{On example models and challenges ahead for the
  evaluation of complex cyber-physical systems with state of the art formal
  methods v\&v}. In \bibinfo{booktitle}{\emph{Air Force Research Laboratory
  Safe and Secure Systems and Software Symposium (S5) Conference, Dayton, OH}}.
  \bibinfo{pages}{9--11}.
\newblock


\bibitem[Ernst et~al\mbox{.}(2022)]{DBLP:conf/arch/ErnstABCDFFG0KM21}
\bibfield{author}{\bibinfo{person}{Gidon Ernst}, \bibinfo{person}{Paolo
  Arcaini}, \bibinfo{person}{Georgios Fainekos}, \bibinfo{person}{Federico
  Formica}, \bibinfo{person}{Jun Inoue}, \bibinfo{person}{Tanmay Khandait},
  \bibinfo{person}{Mohammad~Mahdi Mahboob}, \bibinfo{person}{Claudio Menghi},
  \bibinfo{person}{Giulia Pedrielli}, \bibinfo{person}{Masaki Waga},
  \bibinfo{person}{Yoriyuki Yamagata}, {and} \bibinfo{person}{Zhenya Zhang}.}
  \bibinfo{year}{2022}\natexlab{}.
\newblock \showarticletitle{{ARCH-COMP 2022 Category Report: Falsification with
  Ubounded Resources}}. In \bibinfo{booktitle}{\emph{International Workshop on
  Applied Verification of Continuous and Hybrid Systems (ARCH22)}}
  \emph{(\bibinfo{series}{EPiC Series in Computing},
  Vol.~\bibinfo{volume}{90})}. \bibinfo{publisher}{EasyChair},
  \bibinfo{pages}{204--221}.
\newblock


\bibitem[Formica et~al\mbox{.}(2024a)]{HecateGitHub}
\bibfield{author}{\bibinfo{person}{Federico Formica}, \bibinfo{person}{Tony
  Fan}, \bibinfo{person}{Chris George}, \bibinfo{person}{Alessandro Fischetti},
  {and} \bibinfo{person}{Claudio Menghi}.} \bibinfo{year}{2024}\natexlab{a}.
\newblock \bibinfo{title}{Hecate on GitHub}.
\newblock
\urldef\tempurl \url{https://github.com/Hecate-SBST/Hecate}
\showURL{\tempurl}


\bibitem[Formica et~al\mbox{.}(2024b)]{HecateMatlab}
\bibfield{author}{\bibinfo{person}{Federico Formica}, \bibinfo{person}{Tony
  Fan}, \bibinfo{person}{Chris George}, \bibinfo{person}{Alessandro Fischetti},
  {and} \bibinfo{person}{Claudio Menghi}.} \bibinfo{year}{2024}\natexlab{b}.
\newblock \bibinfo{title}{Hecate on Matlab File Exchange}.
\newblock
\urldef\tempurl \url{https://www.mathworks.com/matlabcentral/fileexchange/173830-hecate}
\showURL{\tempurl}


\bibitem[Formica et~al\mbox{.}(2023a)]{ATheNA}
\bibfield{author}{\bibinfo{person}{Federico Formica}, \bibinfo{person}{Tony
  Fan}, {and} \bibinfo{person}{Claudio Menghi}.}
  \bibinfo{year}{2023}\natexlab{a}.
\newblock \showarticletitle{Search-Based Software Testing Driven by
  Automatically Generated and Manually Defined Fitness Functions}.
\newblock \bibinfo{journal}{\emph{Transactions on Software Engineering and
  Methodologies}} \bibinfo{volume}{33}, \bibinfo{number}{2}, Article
  \bibinfo{articleno}{40} (\bibinfo{year}{2023}), \bibinfo{numpages}{37}~pages.
\newblock
\href{https://doi.org/10.1145/3624745}{doi:\nolinkurl{10.1145/3624745}}


\bibitem[Formica et~al\mbox{.}(2024c)]{Hecate}
\bibfield{author}{\bibinfo{person}{Federico Formica}, \bibinfo{person}{Tony
  Fan}, \bibinfo{person}{Akshay Rajhans}, \bibinfo{person}{Vera Pantelic},
  \bibinfo{person}{Mark Lawford}, {and} \bibinfo{person}{Claudio Menghi}.}
  \bibinfo{year}{2024}\natexlab{c}.
\newblock \showarticletitle{Simulation-Based Testing of Simulink Models With
  Test Sequence and Test Assessment Blocks}.
\newblock \bibinfo{journal}{\emph{IEEE Transactions on Software Engineering}}
  \bibinfo{volume}{50}, \bibinfo{number}{2} (\bibinfo{year}{2024}),
  \bibinfo{pages}{239--257}.
\newblock
\href{https://doi.org/10.1109/TSE.2023.3343753}{doi:\nolinkurl{10.1109/TSE.2023.3343753}}


\bibitem[Formica et~al\mbox{.}(2025)]{RequirementsTables}
\bibfield{author}{\bibinfo{person}{Federico Formica}, \bibinfo{person}{Chris
  George}, \bibinfo{person}{Shayda Rahmatyan}, \bibinfo{person}{Vera Pantelic},
  \bibinfo{person}{Mark Lawford}, \bibinfo{person}{Angelo Gargantini}, {and}
  \bibinfo{person}{Claudio Menghi}.} \bibinfo{year}{2025}\natexlab{}.
\newblock \bibinfo{title}{Search-based Testing of Simulink Models with
  Requirements Tables}.
\newblock
\urldef\tempurl \url{https://arxiv.org/abs/2501.05412}
\showURL{\tempurl}


\bibitem[Formica et~al\mbox{.}(2024f)]{ATheNA_Tool}
\bibfield{author}{\bibinfo{person}{Federico Formica},
  \bibinfo{person}{Mohammad~Mahdi Mahboob}, \bibinfo{person}{Mehrnoosh
  Askarpour}, {and} \bibinfo{person}{Claudio Menghi}.}
  \bibinfo{year}{2024}\natexlab{f}.
\newblock \showarticletitle{ATheNA-S: A Testing Tool for Simulink Models Driven
  by Software Requirements and Domain Expertise}. In
  \bibinfo{booktitle}{\emph{International Conference on the Foundations of
  Software Engineering --- Demonstration}} \emph{(\bibinfo{series}{FSE})}.
  \bibinfo{publisher}{ACM}, \bibinfo{pages}{587--591}.
\newblock


\bibitem[Formica et~al\mbox{.}(2024d)]{ATheNAGitHub}
\bibfield{author}{\bibinfo{person}{Federico Formica},
  \bibinfo{person}{Mohammad~Mahdi Mahboob}, {and} \bibinfo{person}{Claudio
  Menghi}.} \bibinfo{year}{2024}\natexlab{d}.
\newblock \bibinfo{title}{ATheNA on GitHub}.
\newblock
\urldef\tempurl \url{https://github.com/ATheNA-SBST/ATheNA}
\showURL{\tempurl}


\bibitem[Formica et~al\mbox{.}(2024e)]{ATheNAMatlab}
\bibfield{author}{\bibinfo{person}{Federico Formica},
  \bibinfo{person}{Mohammad~Mahdi Mahboob}, {and} \bibinfo{person}{Claudio
  Menghi}.} \bibinfo{year}{2024}\natexlab{e}.
\newblock \bibinfo{title}{ATheNA on Matlab File Exchange}.
\newblock
\urldef\tempurl \url{https://www.mathworks.com/matlabcentral/fileexchange/116095-athena}
\showURL{\tempurl}


\bibitem[Formica et~al\mbox{.}(2023b)]{Hecate_CruiseControl}
\bibfield{author}{\bibinfo{person}{Federico Formica}, \bibinfo{person}{Nicholas
  Petrunti}, \bibinfo{person}{Lucas Bruck}, \bibinfo{person}{Vera Pantelic},
  \bibinfo{person}{Mark Lawford}, {and} \bibinfo{person}{Claudio Menghi}.}
  \bibinfo{year}{2023}\natexlab{b}.
\newblock \showarticletitle{Test Case Generation for Drivability Requirements
  of an Automotive Cruise Controller: An Experience with an Industrial
  Simulator}. In \bibinfo{booktitle}{\emph{European Software Engineering
  Conference and Symposium on the Foundations of Software Engineering}}
  \emph{(\bibinfo{series}{ESEC/FSE})}. \bibinfo{publisher}{ACM},
  \bibinfo{pages}{1949--1960}.
\newblock


\bibitem[Gambi et~al\mbox{.}(2022)]{Gambi_2022_Generating}
\bibfield{author}{\bibinfo{person}{Alessio Gambi}, \bibinfo{person}{Vuong
  Nguyen}, \bibinfo{person}{Jasim Ahmed}, {and} \bibinfo{person}{Gordon
  Fraser}.} \bibinfo{year}{2022}\natexlab{}.
\newblock \showarticletitle{Generating Critical Driving Scenarios from Accident
  Sketches}. In \bibinfo{booktitle}{\emph{2022 IEEE International Conference On
  Artificial Intelligence Testing (AITest)}}. \bibinfo{publisher}{IEEE},
  \bibinfo{pages}{95--102}.
\newblock
\href{https://doi.org/10.1109/AITest55621.2022.00022}{doi:\nolinkurl{10.1109/AITest55621.2022.00022}}


\bibitem[Jin et~al\mbox{.}(2014)]{jin2014powertrain}
\bibfield{author}{\bibinfo{person}{Xiaoqing Jin}, \bibinfo{person}{Jyotirmoy~V
  Deshmukh}, \bibinfo{person}{James Kapinski}, \bibinfo{person}{Koichi Ueda},
  {and} \bibinfo{person}{Ken Butts}.} \bibinfo{year}{2014}\natexlab{}.
\newblock \showarticletitle{Powertrain control verification benchmark}. In
  \bibinfo{booktitle}{\emph{International conference on Hybrid systems:
  computation and control}}. \bibinfo{publisher}{{ACM}},
  \bibinfo{pages}{253--262}.
\newblock


\bibitem[Khandait et~al\mbox{.}(2024)]{Arch2024}
\bibfield{author}{\bibinfo{person}{Tanmay Khandait}, \bibinfo{person}{Federico
  Formica}, \bibinfo{person}{Paolo Arcaini}, \bibinfo{person}{Surdeep
  Chotaliya}, \bibinfo{person}{Georgios Fainekos}, \bibinfo{person}{Abdelrahman
  Hekal}, \bibinfo{person}{Atanu Kundu}, \bibinfo{person}{Ethan Lew},
  \bibinfo{person}{Michele Loreti}, \bibinfo{person}{Claudio Menghi},
  \bibinfo{person}{Laura Nenzi}, \bibinfo{person}{Giulia Pedrielli},
  \bibinfo{person}{Jarkko Peltom\"aki}, \bibinfo{person}{Ivan Porres},
  \bibinfo{person}{Rajarshi Ray}, \bibinfo{person}{Valentin Soloviev},
  \bibinfo{person}{Ennio Visconti}, \bibinfo{person}{Masaki Waga}, {and}
  \bibinfo{person}{Zhenya Zhang}.} \bibinfo{year}{2024}\natexlab{}.
\newblock \showarticletitle{ARCH-COMP 2024 Category Report: Falsification}. In
  \bibinfo{booktitle}{\emph{International Workshop on Applied Verification of
  Continuous and Hybrid Systems. ARCH24}}, Vol.~\bibinfo{volume}{103}.
  \bibinfo{publisher}{EasyChair}, \bibinfo{pages}{122--144}.
\newblock


\bibitem[Khandait et~al\mbox{.}(2025)]{Arch2025}
\bibfield{author}{\bibinfo{person}{Tanmay Khandait}, \bibinfo{person}{Danmay
  Lyu}, \bibinfo{person}{Paolo Arcaini}, \bibinfo{person}{Georgios Fainekos},
  \bibinfo{person}{Federico Formica}, \bibinfo{person}{Sauvik Gon},
  \bibinfo{person}{Abdelrahman Hekal}, \bibinfo{person}{Atanu Kundu},
  \bibinfo{person}{Claudio Menghi}, \bibinfo{person}{Giulia Pedrielli},
  \bibinfo{person}{}, \bibinfo{person}{Rajarshi Ray}, \bibinfo{person}{Quinn
  Thibeault}, \bibinfo{person}{Masaki Waga}, {and} \bibinfo{person}{Zhenya
  Zhang}.} \bibinfo{year}{2025}\natexlab{}.
\newblock \showarticletitle{ARCH-COMP 2025 Category Report: Falsification}. In
  \bibinfo{booktitle}{\emph{International Workshop on Applied Verification of
  Continuous and Hybrid Systems. ARCH25}}. \bibinfo{publisher}{EasyChair}.
\newblock


\bibitem[Khatiri et~al\mbox{.}(2023)]{Khatiri_2023_Simulation}
\bibfield{author}{\bibinfo{person}{Sajad Khatiri}, \bibinfo{person}{Sebastiano
  Panichella}, {and} \bibinfo{person}{Paolo Tonella}.}
  \bibinfo{year}{2023}\natexlab{}.
\newblock \showarticletitle{Simulation-based Test Case Generation for Unmanned
  Aerial Vehicles in the Neighborhood of Real Flights}. In
  \bibinfo{booktitle}{\emph{Conference on Software Testing, Verification and
  Validation}}. \bibinfo{publisher}{IEEE}, \bibinfo{pages}{281--292}.
\newblock
\href{https://doi.org/10.1109/ICST57152.2023.00034}{doi:\nolinkurl{10.1109/ICST57152.2023.00034}}


\bibitem[Majikes et~al\mbox{.}(2013)]{majikes2013literature}
\bibfield{author}{\bibinfo{person}{John~J. Majikes}, \bibinfo{person}{Rahul
  Pandita}, {and} \bibinfo{person}{Tao Xie}.} \bibinfo{year}{2013}\natexlab{}.
\newblock \showarticletitle{Literature review of testing techniques for medical
  device software}. In \bibinfo{booktitle}{\emph{Medical Cyber-Physical Systems
  Workshop}}. \bibinfo{numpages}{8}~pages.
\newblock


\bibitem[Marzella et~al\mbox{.}(2025)]{marzella2025test}
\bibfield{author}{\bibinfo{person}{Michael Marzella}, \bibinfo{person}{Andrea
  Bombarda}, \bibinfo{person}{Marcello Minervini},
  \bibinfo{person}{Nunzio~Marco Bisceglia}, \bibinfo{person}{Angelo
  Gargantini}, {and} \bibinfo{person}{Claudio Menghi}.}
  \bibinfo{year}{2025}\natexlab{}.
\newblock \bibinfo{title}{Test Case Generation for Simulink Models: An
  Experience from the E-Bike Domain}.
\newblock
\urldef\tempurl \url{https://arxiv.org/abs/2501.05792}
\showURL{\tempurl}


\bibitem[MathWorks(2022a)]{benchmarkST}
\bibfield{author}{\bibinfo{person}{MathWorks}.}
  \bibinfo{year}{2022}\natexlab{a}.
\newblock \bibinfo{title}{{Assess a Model by Using When Decomposition}}.
\newblock
  \bibinfo{howpublished}{\url{http://www.mathworks.com/help/sltest/ug/using-when-decomposition-to-write-tests.html}}.
\newblock


\bibitem[MathWorks(2022b)]{benchmarkFS}
\bibfield{author}{\bibinfo{person}{MathWorks}.}
  \bibinfo{year}{2022}\natexlab{b}.
\newblock \bibinfo{title}{{Assess the Damping Ratio of a Flutter Suppression
  System}}.
\newblock
  \bibinfo{howpublished}{\url{http://www.mathworks.com/help/sltest/ug/assess-damping-ratio-of-flutter-suppression-system.html}}.
\newblock


\bibitem[MathWorks(2022c)]{TestSequence}
\bibfield{author}{\bibinfo{person}{MathWorks}.}
  \bibinfo{year}{2022}\natexlab{c}.
\newblock \bibinfo{title}{{Test Sequence}}.
\newblock \bibinfo{howpublished}{Release R2022a. \\
  \url{https://www.mathworks.com/help/sltest/ref/testsequence.html}}.
\newblock


\bibitem[MathWorks(2022d)]{benchmarkTL}
\bibfield{author}{\bibinfo{person}{MathWorks}.}
  \bibinfo{year}{2022}\natexlab{d}.
\newblock \bibinfo{title}{{Test Traffic Light Control by Using Logical and
  Temporal Assessments}}.
\newblock
  \bibinfo{howpublished}{\url{http://www.mathworks.com/help/sltest/ug/test-traffic-light-using-logical-and-temporal-assessments.html}}.
\newblock


\bibitem[MathWorks(2022e)]{benchmarkHPS}
\bibfield{author}{\bibinfo{person}{MathWorks}.}
  \bibinfo{year}{2022}\natexlab{e}.
\newblock \bibinfo{title}{{Use Test Sequence Scenarios in the Test Sequence
  Editor and Test Manager}}.
\newblock
  \bibinfo{howpublished}{\url{http://www.mathworks.com/help/sltest/ug/define-test-sequence-scenarios-in-test-sequence-editor.html}}.
\newblock


\bibitem[Matinnejad et~al\mbox{.}(2015)]{matinnejad2015search}
\bibfield{author}{\bibinfo{person}{Reza Matinnejad}, \bibinfo{person}{Shiva
  Nejati}, \bibinfo{person}{Lionel Briand}, \bibinfo{person}{Thomas Bruckmann},
  {and} \bibinfo{person}{Claude Poull}.} \bibinfo{year}{2015}\natexlab{}.
\newblock \showarticletitle{Search-based automated testing of continuous
  controllers: Framework, tool support, and case studies}.
\newblock \bibinfo{journal}{\emph{Information and Software Technology}}
  \bibinfo{volume}{57} (\bibinfo{year}{2015}), \bibinfo{pages}{705--722}.
\newblock
Issue January 2015.
\href{https://doi.org/10.1016/j.infsof.2014.05.007}{doi:\nolinkurl{10.1016/j.infsof.2014.05.007}}


\bibitem[Mavridou et~al\mbox{.}(2020)]{mavridou2020ten}
\bibfield{author}{\bibinfo{person}{Anastasia Mavridou}, \bibinfo{person}{Hamza
  Bourbouh}, \bibinfo{person}{Dimitra Giannakopoulou}, \bibinfo{person}{Thomas
  Pressburger}, \bibinfo{person}{Mohammad Hejase}, \bibinfo{person}{Pierre-Loic
  Garoche}, {and} \bibinfo{person}{Johann Schumann}.}
  \bibinfo{year}{2020}\natexlab{}.
\newblock \showarticletitle{The ten lockheed martin cyber-physical challenges:
  formalized, analyzed, and explained}. In
  \bibinfo{booktitle}{\emph{International Requirements Engineering
  Conference}}. \bibinfo{publisher}{IEEE}, \bibinfo{pages}{300--310}.
\newblock


\bibitem[Menghi et~al\mbox{.}(2023)]{Arch2023}
\bibfield{author}{\bibinfo{person}{Claudio Menghi}, \bibinfo{person}{Paolo
  Arcaini}, \bibinfo{person}{Walstan Baptista}, \bibinfo{person}{Gidon Ernst},
  \bibinfo{person}{Georgios Fainekos}, \bibinfo{person}{Federico Formica},
  \bibinfo{person}{Sauvik Gon}, \bibinfo{person}{Tanmay Khandait},
  \bibinfo{person}{Atanu Kundu}, \bibinfo{person}{Giulia Pedrielli},
  \bibinfo{person}{Jarkko Peltom\"aki}, \bibinfo{person}{Ivan Porres},
  \bibinfo{person}{Rajarshi Ray}, \bibinfo{person}{Masaki Waga}, {and}
  \bibinfo{person}{Zhenya Zhang}.} \bibinfo{year}{2023}\natexlab{}.
\newblock \showarticletitle{ARCH-COMP 2023 Category Report: Falsification}. In
  \bibinfo{booktitle}{\emph{International Workshop on Applied Verification of
  Continuous and Hybrid Systems. ARCH23}}, Vol.~\bibinfo{volume}{96}.
  \bibinfo{publisher}{EasyChair}, \bibinfo{pages}{151--169}.
\newblock


\bibitem[Menghi et~al\mbox{.}(2025)]{menghi2024completeness}
\bibfield{author}{\bibinfo{person}{Claudio Menghi}, \bibinfo{person}{Eugene
  Balai}, \bibinfo{person}{Darren Valovcin}, \bibinfo{person}{Christoph
  Sticksel}, {and} \bibinfo{person}{Akshay Rajhans}.}
  \bibinfo{year}{2025}\natexlab{}.
\newblock \showarticletitle{Completeness and Consistency of Tabular
  Requirements: an SMT-Based Verification Approach}.
\newblock \bibinfo{journal}{\emph{IEEE Transactions on Software Engineering
  (TSE)}} \bibinfo{volume}{51}, \bibinfo{number}{2} (\bibinfo{year}{2025}),
  \bibinfo{pages}{595--620}.
\newblock


\bibitem[Menghi et~al\mbox{.}(2020)]{Aristeo}
\bibfield{author}{\bibinfo{person}{Claudio Menghi}, \bibinfo{person}{Shiva
  Nejati}, \bibinfo{person}{Lionel Briand}, {and} \bibinfo{person}{Yago~Isasi
  Parache}.} \bibinfo{year}{2020}\natexlab{}.
\newblock \showarticletitle{Approximation-Refinement Testing of
  Compute-Intensive Cyber-Physical Models: An Approach Based on System
  Identification}. In \bibinfo{booktitle}{\emph{International Conference on
  Software Engineering}}. \bibinfo{publisher}{IEEE/ACM},
  \bibinfo{pages}{372--384}.
\newblock


\bibitem[Nejati et~al\mbox{.}(2023)]{Taxonomy}
\bibfield{author}{\bibinfo{person}{Shiva Nejati}, \bibinfo{person}{Lev
  Sorokin}, \bibinfo{person}{Damir Safin}, \bibinfo{person}{Federico Formica},
  \bibinfo{person}{Mohammad~Mahdi Mahboob}, {and} \bibinfo{person}{Claudio
  Menghi}.} \bibinfo{year}{2023}\natexlab{}.
\newblock \showarticletitle{Reflections on Surrogate-Assisted Search-Based
  Testing: A Taxonomy and Two Replication Studies based on Industrial ADAS and
  Simulink Models}.
\newblock \bibinfo{journal}{\emph{Information and Software Technology}}
  \bibinfo{volume}{163} (\bibinfo{year}{2023}), \bibinfo{pages}{107286}.
\newblock
\href{https://doi.org/10.1016/j.infsof.2023.107286}{doi:\nolinkurl{10.1016/j.infsof.2023.107286}}


\bibitem[Papadakis et~al\mbox{.}(2019)]{papadakis2019mutation}
\bibfield{author}{\bibinfo{person}{Mike Papadakis}, \bibinfo{person}{Marinos
  Kintis}, \bibinfo{person}{Jie Zhang}, \bibinfo{person}{Yue Jia},
  \bibinfo{person}{Yves Le~Traon}, {and} \bibinfo{person}{Mark Harman}.}
  \bibinfo{year}{2019}\natexlab{}.
\newblock \showarticletitle{Mutation testing advances: an analysis and survey}.
\newblock In \bibinfo{booktitle}{\emph{Advances in Computers}}.
  Vol.~\bibinfo{volume}{112}. \bibinfo{publisher}{Elsevier},
  \bibinfo{pages}{275--378}.
\newblock


\bibitem[Xue et~al\mbox{.}(2024)]{xue2024domain}
\bibfield{author}{\bibinfo{person}{Zhiyi Xue}, \bibinfo{person}{Liangguo Li},
  \bibinfo{person}{Senyue Tian}, \bibinfo{person}{Xiaohong Chen},
  \bibinfo{person}{Pingping Li}, \bibinfo{person}{Liangyu Chen},
  \bibinfo{person}{Tingting Jiang}, {and} \bibinfo{person}{Min Zhang}.}
  \bibinfo{year}{2024}\natexlab{}.
\newblock \showarticletitle{Domain knowledge is all you need: A field
  deployment of llm-powered test case generation in fintech domain}. In
  \bibinfo{booktitle}{\emph{International Conference on Software Engineering:
  Companion Proceedings}}. \bibinfo{publisher}{IEEE/ACM},
  \bibinfo{pages}{314--315}.
\newblock


\end{thebibliography}

\end{document}